\documentclass[aps,twocolumn,showpacs,preprintnumbers,nofootinbib,prd,superscriptaddress,groupedaddress,10pt]{revtex4-1}

\makeatletter
\def\l@subsubsection#1#2{}
\def\l@subsubsubsection#1#2{}
\makeatother

\usepackage{subfigure}
\usepackage{graphicx,amssymb,amsmath,amsthm,amsfonts,epsfig,epsf}
\usepackage[usenames]{color}
\usepackage{epstopdf}
\definecolor{darkred}{rgb}{0.5,0,0}
\usepackage{latexsym}
\usepackage{array}
\usepackage{afterpage}
\usepackage{bm}
\usepackage{dcolumn}
\usepackage[utf8]{inputenc}
\usepackage{rotating}
\usepackage{longtable}

\usepackage{enumerate}
\usepackage{tensor,multirow}
\usepackage{url}
\usepackage{placeins}
\usepackage[linktocpage]{hyperref}
\usepackage{float}
\usepackage{graphicx}

\def\be{\begin{equation}}
\def\ee{\end{equation}}
\newcommand{\beq}{\begin{eqnarray}}
\newcommand{\eeq}{\end{eqnarray}}

\def\ba{\begin{align}}
\def\ea{\end{align}}

\newcounter{mnotecount}[section]

\renewcommand{\themnotecount}{\thesection.\arabic{mnotecount}}

\newcommand{\mnote}[1]
{\protect{\stepcounter{mnotecount}}$^{\mbox{\footnotesize
$
\bullet$\themnotecount}}$ \marginpar{
\raggedright\tiny\em
$\!\!\!\!\!\!\,\bullet$\themnotecount: #1} }

\begin{document}

\title{Charged Fermions and Strong Cosmic Censorship}

\author{
Kyriakos Destounis
}
\affiliation{ CENTRA, Departamento de F\'{\i}sica, Instituto Superior T\'ecnico -- IST, Universidade de Lisboa -- UL,
Avenida Rovisco Pais 1, 1049 Lisboa, Portugal}
\begin{abstract}
It was recently shown that the Strong Cosmic Censorship conjecture might be violated for near-extremally-charged black holes in de Sitter space. Here, we extend our study to charged fermionic fields in the exterior of Reissner-Nordstr\"{o}m-de Sitter black holes. We identify three families of modes; one related to the photon sphere, a second related to the de Sitter horizon and a third which dominates near extremality. We show that for near-extremally-charged black holes there is a critical fermionic charge below which Strong Cosmic Censorship may potentially be violated. Surprisingly enough, as one approaches extremality even more, violation of Strong Cosmic Censorship may occur even beyond the critical fermionic charge. 
\end{abstract}

\maketitle
%

\noindent{\bf{\em I. Introduction.}}
We recently studied the implications of massless neutral scalar perturbations on Strong Cosmic Censorship (SCC) in Reissner--Nordstr\"{o}m (RN) black holes (BHs) in de Sitter (dS) spacetime \cite{Cardoso:2017soq}. Three different families of modes were identified in such spacetime; one directly related to the photon sphere, well described by standard  Wentzel-Kramers-Brillouin (WKB) tools, another family whose existence and timescale is closely related to the dS horizon and a third family which dominates for near-extremally-charged BHs. Surprisingly enough, our results show that near-extremal RNdS BHs might violate SCC at the linearized level, leading to a possible failure of determinism in General Relativity (GR). The key quantity controlling the stability of the Cauchy horizon (CH), and therefore the fate of SCC, is given by~\cite{Hintz:2015jkj,CostaFranzen}
\begin{equation}
\label{betaOld}
\beta \equiv -\text{Im}(\omega_0)/\kappa_-\;,
\end{equation}
where $\omega_0$ is the longest-lived/dominant non-zero quasinormal
mode (QNM~\cite{Kokkotas:1999bd,Berti:2009kk,Konoplya:2011qq}) and $\kappa_-$ is the surface gravity of the CH.
The results in~\cite{Hintz:2016gwb,Hintz:2016jak,CGNS4} suggest that $\beta$ remains the key quantity in the non-linear setting: the higher $\beta$, the more stable the CH. Concretely, the modern formulation of SCC requires that
\begin{equation}\label{theoremHintz}
\beta < 1/2
\end{equation}
in order to guarantee the breakdown of field equations at the CH \cite{Christodoulou:2008nj}.
In~\cite{Cardoso:2017soq}, a thorough linear numerical study of $\beta$ throughout the whole parameter space of subextremal RNdS spacetimes revealed that $1/2<\beta<1$ in the near-extremal regime, leading to a CH with enough regularity for metric extensions to be possible past it. 
This provides evidence for the existence of CHs which, upon perturbation, are rather singular due to the divergence of curvature invariants, but where the gravitational field can still be described by the field equations; the evolution of gravitation beyond the CH, however, is highly non-unique. Recent studies \cite{Liu:2019lon,Rahman:2018oso} have generalized the linear massless scalar field study in higher-dimensional RNdS BHs and RNdS BHs on the brane finding potential violation in the near-extremal regime.

There are different ways to interpret the results of \cite{Cardoso:2017soq}. If one takes the SCC conjecture as a purely mathematical question about GR then the results of \cite{Cardoso:2017soq} either signify a failure of SCC, or are superseded by nonlinear effects. In fact, the results of \cite{Luna:2018jfk} proved that even nonlinear effects could not save the conjecture from failing for near-extremally-charged BHs. 

An interesting suggestion to restore SCC, in the presence of a positive cosmological constant, was proposed in \cite{Dafermos:2018tha}, where it was shown that the pathologies identified in \cite{Cardoso:2017soq} become non-generic if one considerably enlarges the allowed set of initial data by weakening their regularity. The considered data are also compatible with Christodoulou's formulation of SCC.


A subsequent study of metric fluctuations in RNdS BHs showed that such perturbations possibly exhibit a much worse violation of SCC. In \cite{Dias:2018etb} it was shown that for a sufficiently large near-extremal RNdS BH, perturbations arising from smooth initial data can be extended past the CH in an arbitrarily smooth way.
Nevertheless, astrophysical BHs are expected to be nearly neutral~\cite{Barausse:2014tra,Cardoso:2016olt}. Taking this into consideration, one can question the relevance of SCC violations in highly charged, non-spinning BHs. In fact, a recent study suggests that rapidly rotating BHs in cosmological backgrounds do not violate SCC. According to \cite{Dias:2018ynt}, in Kerr-dS spacetime (\ref{betaOld}) remains unchanged, but now $\beta$ seems to be bounded exactly by $1/2$, at extremal rotation. Similar results were obtained in \cite{Rahman:2018oso} for higher-dimensional Kerr-dS BHs.

Considering the formation of a charged BH, one would argue that charged matter has to be present. In \cite{Hod:2018dpx,Hod:2018lmi}, it was claimed that charged scalar fields would lead to restoration of SCC in an appropriate region of the parameter space of RNdS and Kerr-Newmann-dS BHs. This implication requires working in the large-coupling regime for which $\beta<1/2$. Taking into account the whole parameter space, subsequent studies \cite{PhysRevD.98.104007,Dias:2018ufh,Hongbao} presented numerical evidence that SCC may still be violated in the setting of charged scalar perturbations in RNdS. 

In \cite{PhysRevD.98.104007}, massive charged scalars were taken into account providing evidence that $\beta>1$. Recall that this is related to bounded curvature and therefore opens the possibility to the existence of solutions to the Einstein-Maxwell-Klein-Gordon system with a scalar field satisfying Price's law and bounded curvature across the CH. Nonetheless, if the neutral scalar perturbations where superimposed to the charged massive ones, then the smaller of the two types of perturbations is the one relevant for SCC, thus getting $\beta<1$, which should be enough to guarantee the blow-up of curvature components. 

In \cite{Dias:2018ufh} it was shown that even for large scalar field charge there are near-extremal BHs for which $\beta>1/2$. A key ingredient of the aforementioned studies was the existence of a superradiantly unstable mode. In fact, it has been shown that linear instabilities arise in various setups (for an incomplete list see \cite{Brito:2015oca,Destounis:2018utr,Zhu:2014sya,Konoplya:2014lha,Furuhashi:2004jk,Dolan:2012yt}). This unstable mode was the dominant one for small scalar charges rendering the question of the validity of SCC irrelevant for a significant region of the parameter space.

Is it natural to question then, if the charged matter could just as well be fermionic instead of scalar. Fermions do not superradiate, leaving the entire range of fermionic charge open for the study of SCC at the linearized level. 
The results of \cite{Ge:2018vjq} provide evidence that fermionic perturbations of RNdS BHs might violate SCC for sufficiently large BH charge. As a matter of fact, the family that seems to dominate the dynamics near extremality is, mostly, the photon sphere family with a very small participation of a family which is purely imaginary for zero fermionic charge $q$ and quickly becomes subdominant as $q$ increases. Unfortunately, there is no information about the classification of the latter family and if it will eventually dominate the dynamics for even higher BH charges. Moreover, since a dS horizon is present, the dS family of modes might be present as well and even dominate the dynamics for small cosmological constants in analogy with what was found in \cite{Cardoso:2017soq}. The tool used to extract the modes in \cite{Ge:2018vjq} is time domain analysis. Although it is a very powerful tool for such calculations, there is a slight chance that long-lived modes may be missed either because of their timescale being larger than the evolution time of the system or cause of improper initial data.

In this paper, we study the propagation of charged fermions on a fixed RNdS background and extract the QNMs with the spectral method in \cite{Jansen:2017oag} which is based on numerical methods introduced in \cite{Dias:2010eu} (for a topical review see \cite{Dias:2015nua}). After characterizing the families of modes that are present, we will examine the implications on SCC for near-extremal RNdS BHs.
  
\noindent{\bf{\em II. Charged Fermions in Reissner-Nordstr\"{o}m-de Sitter spacetime.}}
We focus on RNdS BHs, described by the metric
\begin{equation}
\label{RNdS_space}
ds^2=-f(r)dt^2+\frac{dr^2}{f(r)}+r^2(d\theta^2+\sin^2\theta d\varphi^2)\,,
\end{equation}
where $f(r)=1-{2M}{r^{-1}}+{Q^2}{r^{-2}}-\Lambda r^2/3$. Here, $M,\,Q$ are the BH mass and charge, respectively, and $\Lambda>0$ is the cosmological constant. The surface gravity of each horizon is then
\begin{equation}
\label{surfGrav}
\kappa_h= \frac{1}{2}|f'(r_h)|\;\;,\; h\in\{-,+,c\}\;,
\end{equation}
where $r_-<\,r_+<\, r_c$ are the Cauchy, event and cosmological horizon radius. Since fermions are described by spinors, we use the tetrad formalism to accommodate them in curved space. The tetrads by definition satisfy the relations
\begin{align*}
e^{(a)}_\mu \,e^{\nu}_{(a)}&=\delta^\nu_\mu,\\
e^{(a)}_\mu \,e^{\mu}_{(b)}&=\delta^{(a)}_{(b)},
\end{align*}
The choice of the tetrad field determines the metric through
\begin{align*}
\label{tetradgmn}
g_{\mu\nu}&=e^{(a)}_\mu \,e^{(b)}_\nu\,\eta_{(a)(b)},\\
\eta_{(a)(b)}&=e^\mu_{(a)}\,e^\nu_{(b)}\,g_{\mu\nu},
\end{align*}
where $\eta_{(a)(b)}$ and $g_{\mu\nu}$ are the Minkowski and RNdS metric, respectively. In order to write the Dirac equation, we also introduce the spacetime-dependent gamma matrices $G^\mu$ which are related to the special relativity matrices, $\gamma^{(a)}$, by
\begin{equation*}
\label{gtet}
G^\mu=e^\mu_{(a)}\gamma^{(a)},
\end{equation*}
and are chosen in a proper way to satisfy the anti-commutation relations
\begin{align*}
\{\gamma^{(a)},\gamma^{(b)}\}&=- 2\eta^{(a)(b)},\\
\{G^\mu, G^\nu\}&=- 2 g^{\mu\nu}.
\end{align*}
Consequently, we define $G^\mu$ with respect to a fixed tetrad
as 
\begin{align*}
G^t&=e^t_{(a)}\gamma^{(a)}=\frac{\gamma^t}{\sqrt{f(r)}},\,\,\,\,\,\,\,\,G^r=e^r_{(a)}\gamma^{(a)}=\sqrt{f(r)}\gamma^r,\\
G^\theta&=e^\theta_{(a)}\gamma^{(a)}=\gamma^\theta,\,\,\,\,\,\,\,\,\,\,\,\,\,\,\,\,\,\,\,G^\varphi=e^\varphi_{(a)}\gamma^{(a)}=\gamma^\varphi,
\end{align*}
where $\gamma^t,\,\gamma^r,\,\gamma^\theta$ and $\gamma^\varphi$ are the $\gamma-$matrices in "polar coordinates" \cite{Finster:1998ak}
\begin{align*}
\gamma^t&=\gamma^{(0)},\\
\gamma^r&=\sin\theta\cos\varphi\,\gamma^{(1)}+\sin\theta\sin\varphi\,\gamma^{(2)}+\cos\theta\,\gamma^{(3)}\\
\gamma^\theta&=\frac{1}{r}\left(\cos\theta\cos\varphi\,\gamma^{(1)}+\cos\theta\sin\varphi\,\gamma^{(2)}-\sin\theta\,\gamma^{(3)}\right),\\
\gamma^\varphi&=\frac{1}{r\,\sin\theta}\left(-\sin\varphi\,\gamma^{(1)}+\cos\varphi\,\gamma^{(2)}\right)
\end{align*}
and 
\begin{equation*}
\label{dirac_gamma}
\gamma^{(0)}=\begin{pmatrix}
\mathbf{1}&0\\
0&-\mathbf{1}
\end{pmatrix},\,\,\,\,\,\,\,\,\,
\gamma^{(k)}=\begin{pmatrix}
0&\sigma^{k}\\
-\sigma^{k}&0
\end{pmatrix}
\end{equation*}
the standard Dirac $\gamma$-matrices, where $\sigma^k,\,k=1,2,3$ the Pauli matrices. The propagation of a spin $1/2$ particle of mass $m_f$ on a fixed RNdS background is then described by the Dirac equation in curved spacetime \cite{Fock:1929vt}
\begin{equation}
\label{dirac equation}
(iG^\mu D_\mu-m_f)\psi=0,
\end{equation}
with the covariant derivative 
\begin{equation*}
D_\mu=\partial_\mu-iqA_\mu+\Gamma_\mu,
\end{equation*}
where $q$ the charge of the Dirac particle, $A=-(Q/r)dt$ the electrostatic potential and $\Gamma_\mu$ the spin connection coefficients defined as
\begin{equation*}
\label{spin}
\Gamma_\mu=-\frac{1}{8}\omega_{(a)(b)\mu}\left[\gamma^{(a)},\gamma^{(b)}\right].
\end{equation*}
The spin connection $\omega_{(a)(b)\mu}$ is defined as
\begin{align*}
\omega_{(a)(b)\mu}=\eta_{(a)(c)}\left(e^{(c)}_\nu e^\lambda_{(b)}\Gamma^\nu_{\mu\lambda}-e^\lambda_{(b)}\partial_\mu e^{(c)}_\lambda\right),
\end{align*}
with $\Gamma^\nu_{\mu\lambda}$ the Christoffel symbols. By choosing the ansatz $\psi=f(r)^{-1/4}r^{-1}\Psi$, (\ref{dirac equation}) can be written as
\begin{align}
\nonumber
\label{dirac_2}
\left[\frac{i\gamma^t}{\sqrt{f(r)}}\frac{\partial}{\partial t}+i\sqrt{f(r)}\gamma^r\frac{\partial}{\partial r}-\frac{i\gamma^r}{r}+i\left(\gamma^\theta\frac{\partial}{\partial\theta}+\gamma^\varphi\frac{\partial}{\partial\varphi}\right)\right.\\
\left.-\gamma^{t}\frac{q Q}{r\sqrt{f(r)}}-m_f\right]\Psi=0.
\end{align}
Since the external fields are spherically symmetric and time-independent, we can separate out the angular and time dependence of the wave functions via spherical harmonics and plane waves, respectively. For the Dirac wavefunctions, we choose the ansatzes
\begin{align}
\label{spinor1}
\Psi^+_{jk\omega}&=e^{-i\omega t}\begin{pmatrix}
\phi^k_{j-1/2}F^+(r)\\
i\phi^k_{j+1/2}G^+(r)
\end{pmatrix},\\
\label{spinor2}
\Psi^-_{jk\omega}&=e^{-i\omega t}\begin{pmatrix}
\phi^k_{j+1/2}F^-(r)\\
i\phi^k_{j-1/2}G^-(r)
\end{pmatrix},
\end{align}
where we introduced the spinor spherical harmonics \cite{Finster:1998ak} 
\begin{align*}
\phi^k_{j-1/2}&=\begin{pmatrix}
\sqrt{\frac{j+k}{2j}}Y^{k-1/2}_{j-1/2}(\theta,\varphi)\\
\sqrt{\frac{j-k}{2j}}Y^{k+1/2}_{j-1/2}(\theta,\varphi)
\end{pmatrix},\,\,\,\,\,\,\,\,\,\,\,\,\,\,\,\,\text{for}\,\,\,\,j=l+\frac{1}{2},\\
\phi^k_{j+1/2}&=\begin{pmatrix}
\sqrt{\frac{j+1-k}{2j+2}}Y^{k-1/2}_{j+1/2}(\theta,\varphi)\\
-\sqrt{\frac{j+1+k}{2j+2}}Y^{k+1/2}_{j+1/2}(\theta,\varphi)
\end{pmatrix},\,\,\,\,\,\,\text{for}\,\,\,\,j=l-\frac{1}{2},
\end{align*}
with $j=1/2,3/2,\dots$, $k=-j,-j+1,\dots,j$ and $Y_l^m$ the ordinary spherical harmonics. By substituting (\ref{spinor1}) and (\ref{spinor2}) into (\ref{dirac_2}) and utilizing the identities
\begin{align*}
K=\vec{\sigma}\vec{L}+\mathbf{1}&=-r\sigma^r\left(\sigma^\theta\partial_\theta+\sigma^\varphi\partial_\varphi\right)+\mathbf{1},\\
K \phi^k_{j\mp 1/2}&=\pm(j+\frac{1}{2})\phi^k_{j\mp 1/2},\\
\sigma^r\phi^k_{j\mp 1/2}&=\phi^k_{j\pm 1/2},
\end{align*}
with $\vec{\sigma},\,\vec{L}$ the Pauli and angular momentum vectors, respectively, and $\sigma^r,\,\sigma^\theta,\,\sigma^\varphi$ the Pauli matrices in "polar coordinates" \cite{Finster:1998ak}, we end up with the coupled Dirac equations
\begin{align}
\label{final1}
\frac{\partial F}{\partial r_*}-\frac{\xi\sqrt{f(r)}}{r}F+\left(\omega -\frac{qQ}{r}\right)G+m_f\sqrt{f(r)}G&=0,\\
\label{final2}
\frac{\partial G}{\partial r_*}+\frac{\xi\sqrt{f(r)}}{r}G-\left(\omega -\frac{qQ}{r}\right)F+m_f\sqrt{f(r)}F&=0,
\end{align}
where $\xi=\pm(j+1/2)=\pm 1,\,\pm 2,\dots$ and $dr_*=f/dr$. Since the charge-to-mass ratio of the electron is of order $10^{11} \text{C}/\text{kg}$, it is reasonable to explore massless fermions. By setting $m_f=0$ we can decouple (\ref{final1}), (\ref{final2}) by introducing a new coordinate
\begin{equation*}
\label{new_tort}
d\bar{r}_*=\frac{\left(1-\frac{qQ}{r\omega}\right)}{f}dr,
\end{equation*}
to get
\begin{align}
\label{almdec1}
\frac{dF}{d\bar{r}_*}-WF+\omega G&=0,\\
\label{almdec2}
\frac{dG}{d\bar{r}_*}+WG-\omega F&=0,
\end{align}
and subsequently
\begin{align}
\frac{d^2F}{d\bar{r}_*^2}+\left(\omega^2-V_+\right)F&=0,\\
\frac{d^2G}{d\bar{r}_*^2}+\left(\omega^2-V_-\right)G&=0,
\end{align}
with 
\begin{equation*}
V_\pm=\pm\frac{dW}{d\bar{r}_*}+W^2,
\end{equation*}
where
\begin{equation}
\label{W}
W=\frac{\xi\sqrt{f}}{r\left(1-\frac{qQ}{r\omega}\right)}.
\end{equation}
It can be shown that potentials related in this manner and subjected to Sommerfeld conditions are isospectral, thus, allowing us to work only with the field $F$~\cite{Berti:2009kk,Anderson:1991kx}. Since we are interested in the characteristic frequencies of this spacetime, we impose the boundary conditions
\begin{equation*}
F(r\rightarrow r_+)\sim e^{-i\omega \bar{r}_*},\,\,\,\,\,\,\,\,\,\,\,\,\,\,\,F(r\rightarrow r_c)\sim e^{i\omega \bar{r}_*}
\end{equation*}
which select a discrete set of frequencies $\omega$ called the QNMs. The QN frequencies are characterized, for each $\xi$, by an integer $n\geq 0$ labeling the mode number. The fundamental mode $n=0$ corresponds, by definition, to the non-vanishing frequency with the smallest (in absolute value) imaginary part and will be denoted by $\omega\neq 0$. It is apparent from (\ref{almdec1}), (\ref{almdec2}) and (\ref{W}) that the symmetry $\omega\rightarrow-\omega$, $q\rightarrow-q$, $\xi\rightarrow-\xi$ holds, enabling us to only study positive $\xi$.
\begin{figure*}[t]
\subfigure{\includegraphics[scale=0.24]{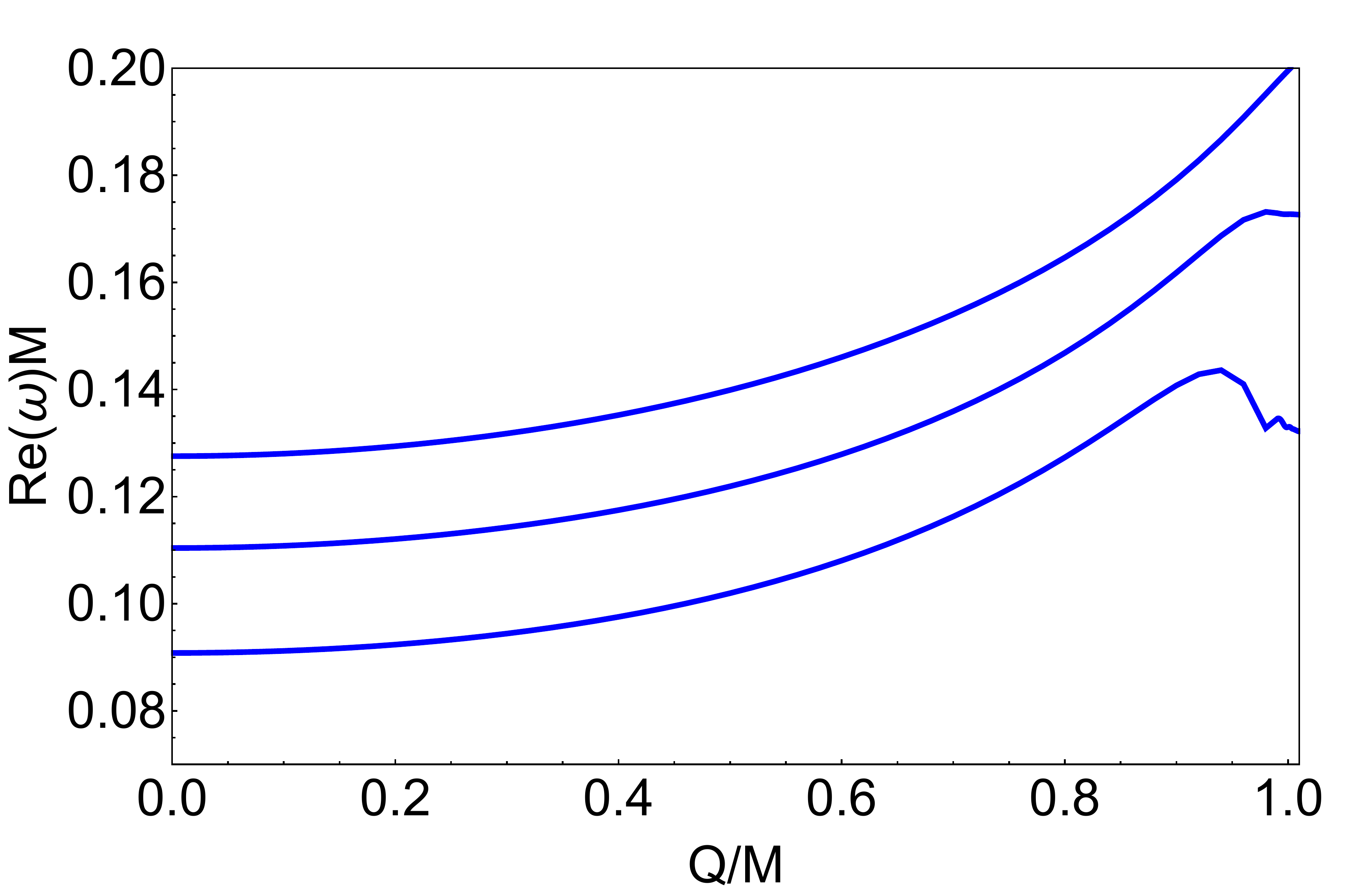}}
\subfigure{\includegraphics[scale=0.24]{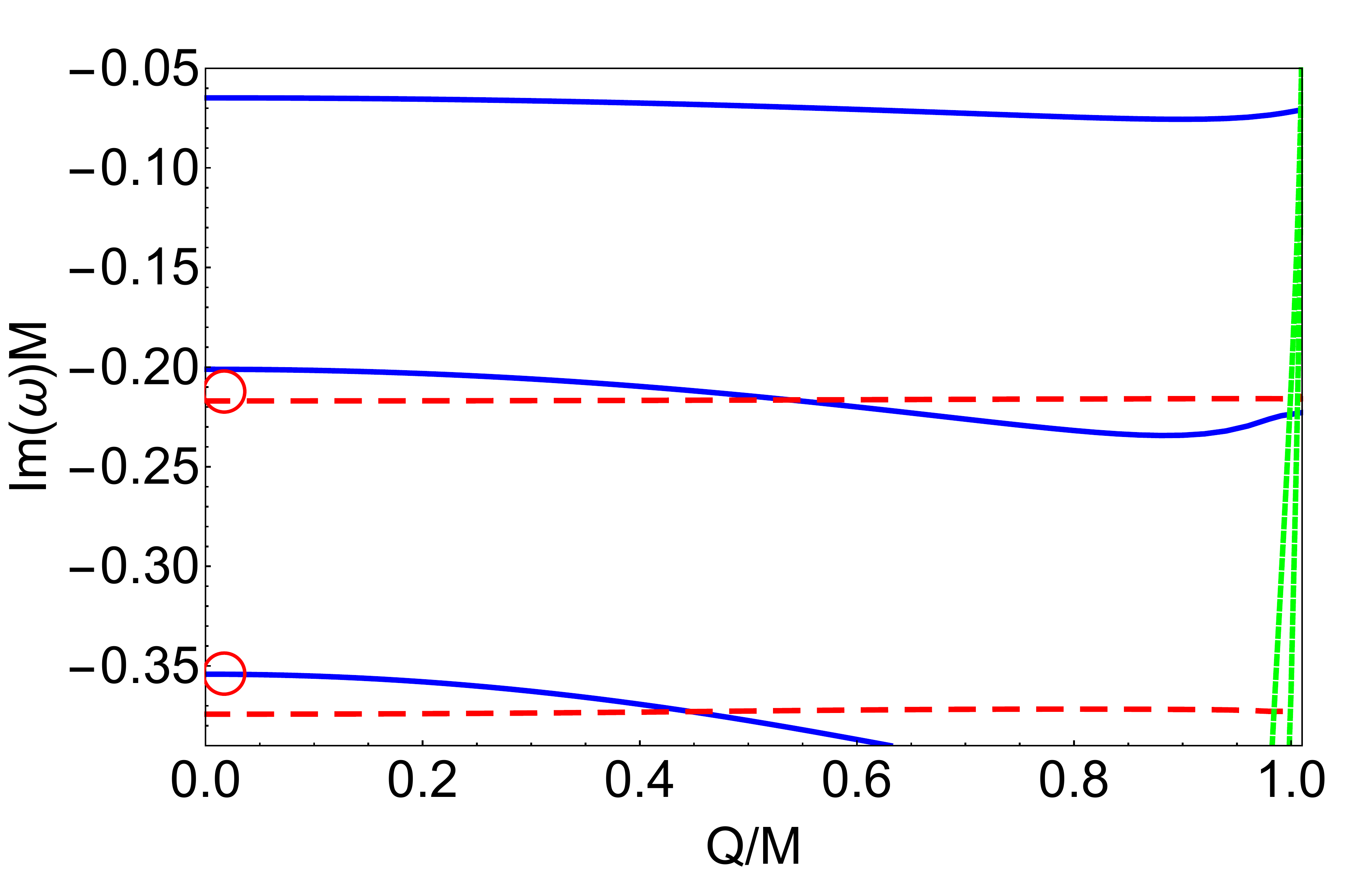}}
\caption{Lowest lying fermionic QNMs for $\xi=1$, $q=0$ and $\Lambda M^2=0.06$ as a function of $Q/M$. The right plot shows the imaginary part, with dashed red lines corresponding to purely imaginary modes, and solid blue lines to complex PS modes, whose real part is shown in the left plot. The red circles in the right plot designate the $\xi=1$ Dirac modes of empty dS space at the same $\Lambda$, which closely match the first imaginary mode shown here, but lie less close to the higher overtone. Near extremality, another set of purely imaginary modes (dotted green lines) come in from $-\infty$ and aprroach $0$ in this limit. Only a finite number of modes are shown, even though we expect infinitely many near-extremal modes in the range shown.}
\label{QNMs}
\end{figure*}
As shown in \ref{app:beta}, for $q\neq 0$ the stability of the CH continues to be determined by~\eqref{betaOld}.

The results shown in the following sections were obtained with the Mathematica package of~\cite{Jansen:2017oag}, and checked in various cases with a WKB approximation~\cite{Iyer:1986np} and with a code developed based on the matrix method \cite{KaiLin1}.
\begin{figure*}[ht!]
\includegraphics[scale=0.24]{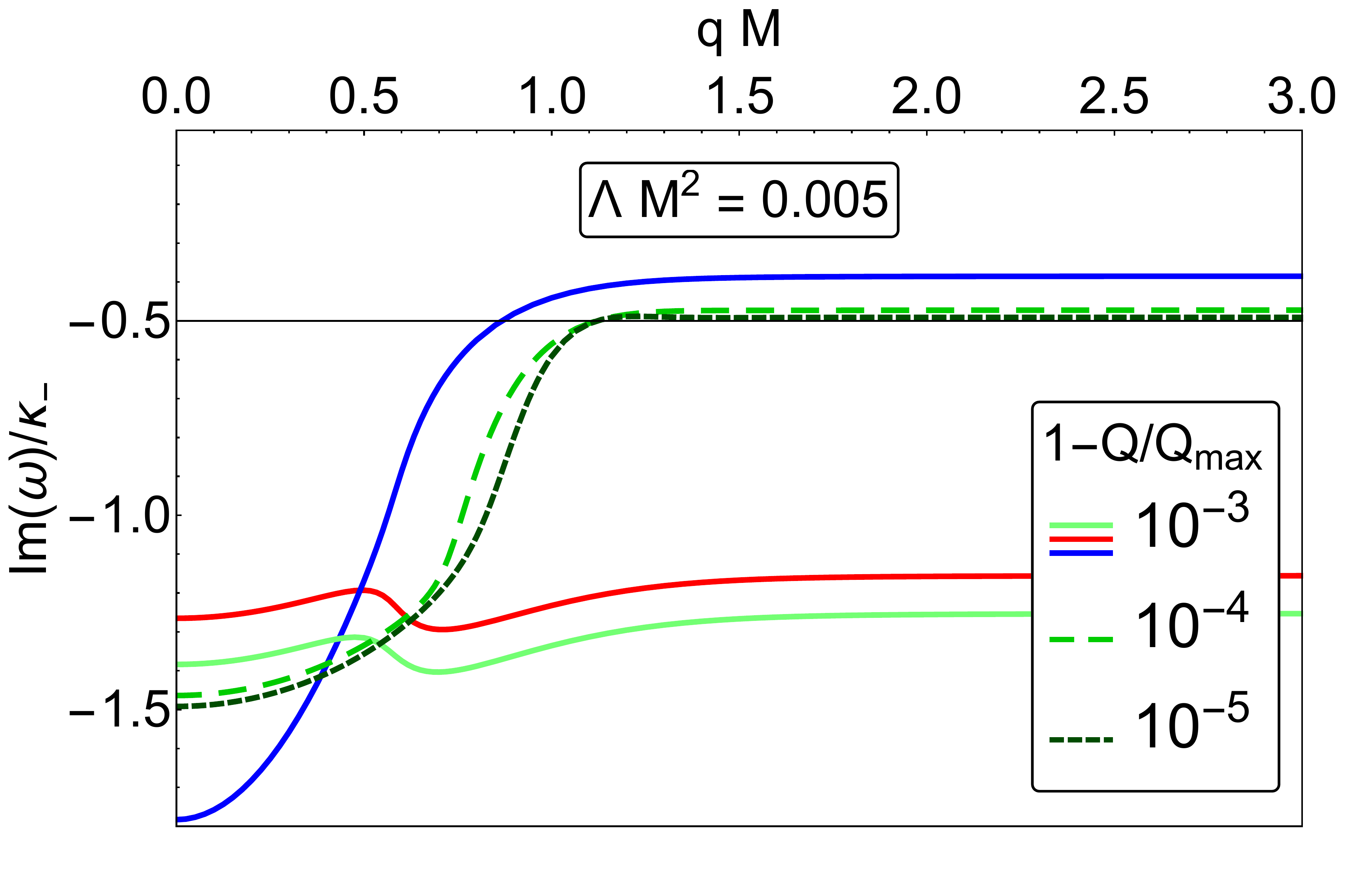}
\includegraphics[scale=0.24]{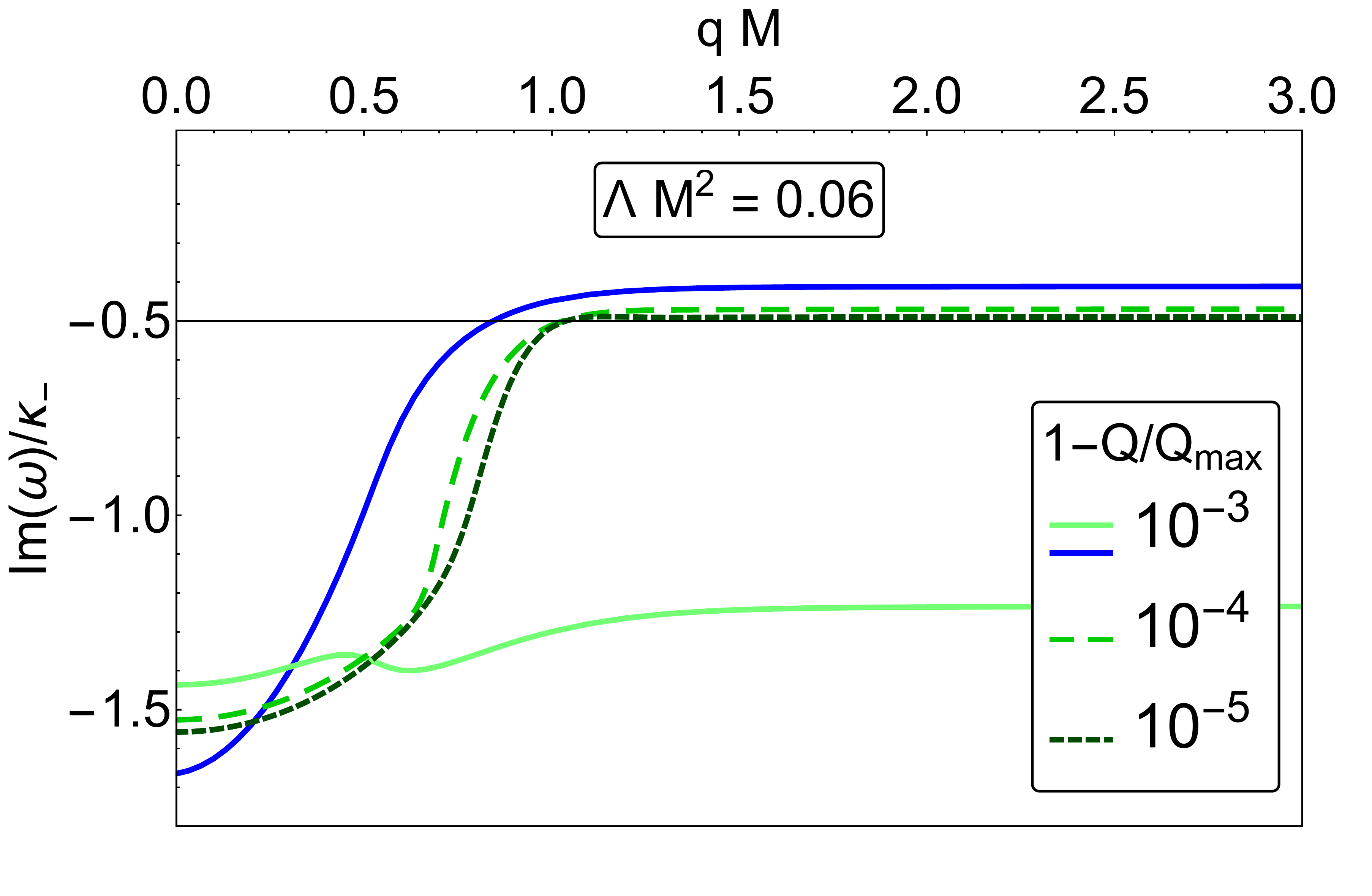}
\caption{Lowest lying QNMs of a charged, massless fermionic perturbation of a RNdS BH with $\xi=1$, $\Lambda M^2=0.005,\,0.06$ and $1-Q/Q_\text{max}=10^{-3},\,10^{-4},\,10^{-5}$ as a function of the fermionic charge $q M$. The modes denoted with blue, red and green colors belong to the PS, dS and NE family, respectively.}
\label{smallL}
\end{figure*}

\noindent{\bf{\em III. QNMs of massless, charged fermionic fields: the three families.}}
In \cite{Cardoso:2017soq,PhysRevD.98.104007},
we found three qualitatively different families of QNMs: the photon sphere (PS) family, the dS family and the near-extremal (NE) family.
The first two connect smoothly to the modes of asymptotically flat Schwarzschild and of empty dS, respectively, while the last family cannot be found in either of these spacetimes.
Here we, again, distinguish three families of modes.

\begin{figure}[t]
\begin{center}
\includegraphics[scale=0.24]{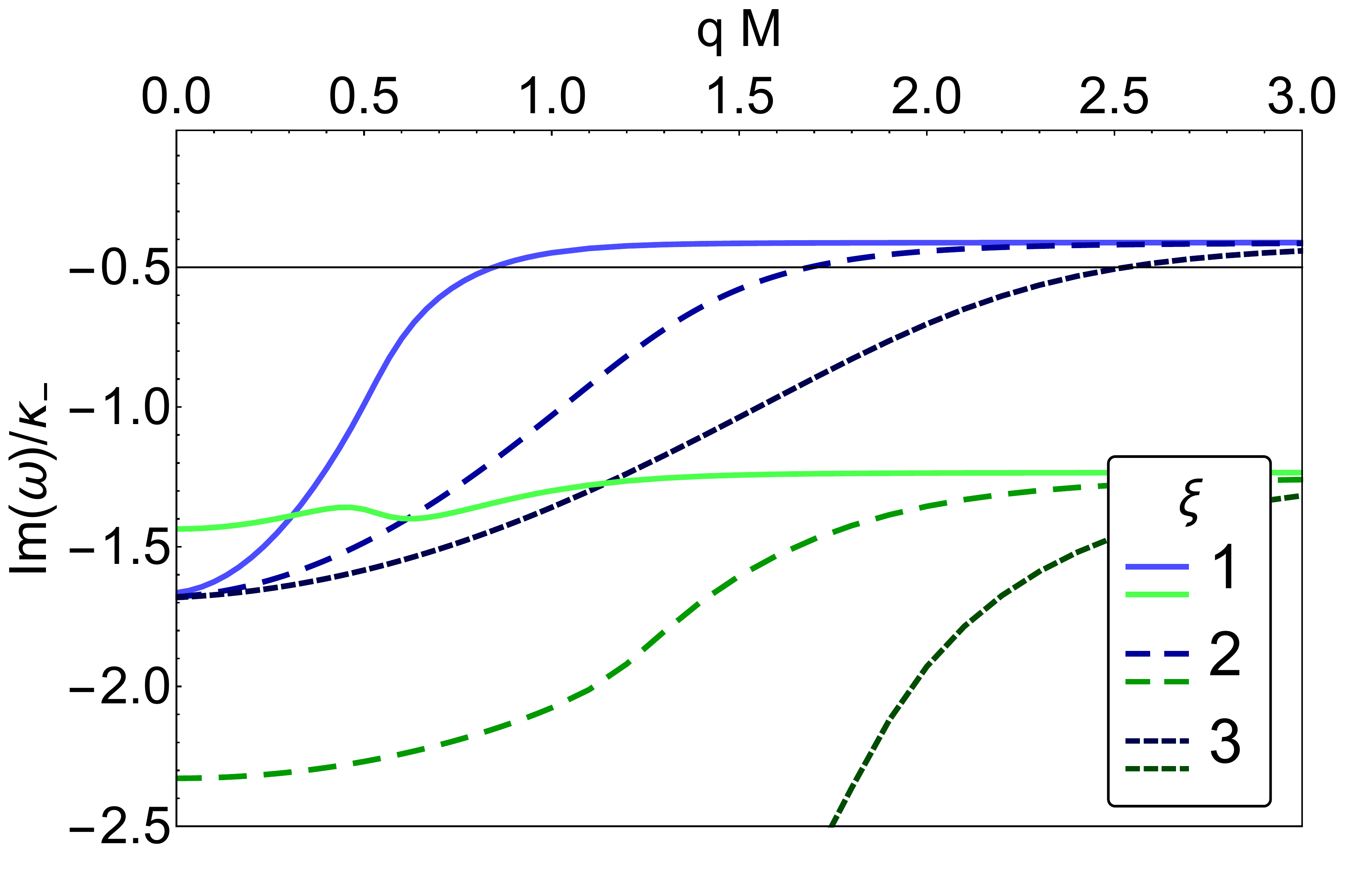}
\end{center}
\caption{Lowest lying QNMs of a charged, massless fermionic perturbation of a RNdS BH for various $\xi$ with $\Lambda M^2=0.06$ and $1-Q/Q_\text{max}=10^{-3}$ as a function of the fermionic charge $qM$.  The modes denoted with blue and green colors belong to the PS and NE family, respectively. The dS family is absent in this range.}
\label{higher_k}
\end{figure}

\noindent{\bf{\em Photon sphere modes.}}
The PS is a spherical trapping region of space where gravity is strong enough that photons are forced to travel in unstable circular orbits around a BH. This region has a strong pull in the control of decay of perturbations and the QNMs with large frequencies. For asymptotically dS BHs, we find a family that can be traced back to the photon sphere and refer to them as PS modes. These modes are shown with blue colors in Figs. \ref{QNMs}-\ref{smallL} and \ref{higher_k}. They satisfy the symmetry $\omega\rightarrow-\omega$ for $q=0$ and the symmetry breaks as the fermionic charge is turned on, according to (\ref{W}). For very small $\Lambda$, $q$ and $Q$, $\xi\rightarrow\infty$ defines the dominant mode which can be very well approximated by a WKB approximation and asymptote to the Schwarzschild BH Dirac QNMs \cite{Cho:2003qe}. The lowest lying PS modes are weekly dependent on the BH charge as it is apparent for the case presented in Fig. \ref{QNMs}. For sufficiently large $\Lambda$ the former does not hold. For large BH charges the $\xi=1$ PS modes dominate the family (see \ref{higher}). As the BH "disappears" ($M\rightarrow 0$) we observe that the PS family have increasingly large frequencies and timescales until they abruptly vanish (see \ref{dS}).

It is important to note that at the eikonal limit the fermionic PS QNMs coincide with the scalar ones. This occurs since at the eikonal limit the effective potentials for fermionic and scalar perturbations are dominated by the angular numbers $\xi$, $l$ thus gaining a similar form. A basic difference between scalar and fermionic perturbations
is that the eikonal scalar QNMs are the dominant ones for all BH parameters, in contrast with the eikonal fermionic QNMs which are dominant for a very small region of the BH parameter space.

\noindent{\bf{\em de Sitter modes.}}
In pure dS space solutions of the Dirac equation with purely imaginary $\omega$ exist \cite{LopezOrtega:2006ig}
\begin{equation}
\label{pure_dS}
\omega_\text{pure dS}/\kappa^\text{dS}_c=-i\left(\xi+n+\frac{1}{2}\right)
\end{equation}
where $\xi=1,2,\dots$. The second family of modes we find are the Dirac BH dS QNMs, which are deformations of pure dS QNMs (\ref{pure_dS}). The dominant BH dS mode ($\xi=1$, $n=0$) has almost identical imaginary part with (\ref{pure_dS}) and higher overtones have increasingly larger deformations.

These modes have weak dependence on the BH charge and are described by the surface gravity $\kappa_c^\text{dS}=\sqrt{\Lambda/3}$ of the cosmological horizon of pure dS space, as opposed to that of the cosmological horizon in the RNdS BH in study. This could be explained by the fact that the accelerated expansion of RNdS spacetimes is also governed by $\kappa_c^\text{dS}$ \cite{Brill:1993tw,Rendall:2003ks}.

To the best of our knowledge, this family of Dirac BH dS modes has been identified here for the first time. The scalar equivalent of these modes has been identified for the first time in the QNM calculations of \cite{Jansen:2017oag, Cardoso:2017soq}. Moreover, as the black hole vanishes ($M \rightarrow 0$), these modes converge smoothly to the exact pure dS modes (\ref{pure_dS}) (see \ref{dS}). 

A key similarity of fermionic dS QNMs and scalar dS QNMs is the fact that they are both
proportional to the surface gravity of the cosmological horizon of pure dS space (see \cite{Cardoso:2017soq}). On
the other hand, the fermionic dS QNMs do not admit an $\omega=0$ mode, while scalars do.
Such mode has been seen in time evolutions \cite{Zhu:2014sya,Konoplya:2014lha} and rises from the fact that the effective
potential forms a potential well right outside the photon sphere, serving as a trapping region.
This region is connected to a superradiant instability in RNdS against charged scalar fluctuations
which effectively puts the validity of SCC out of discussion since the internal and external 
regions of the BH in study are effectively unstable. The effective potential of fermionic 
perturbations does not contain any potential wells which is gonna play an important role in our 
discussion.

\noindent{\bf{\em Near-extremal modes.}}
In the limit where the Cauchy and event horizon approach each other, a third NE family dominates the dynamics. In the extremal limit and for sufficiently small fermionic charges this family approaches
\begin{equation}
\label{near extremal}
\omega_\text{NE}\approx\frac{q Q}{r_-}-i\kappa_-\left(\xi+n+\frac{1}{2}\right)\approx\frac{q Q}{r_+}-i\kappa_+\left(\xi+n+\frac{1}{2}\right),
\end{equation}
where $\xi=1,2,\dots$, with weak dependence on $\Lambda$ as shown by our numerics. As indicated by (\ref{near extremal}), the dominant mode of this family is the $\xi=1$, $n=0$. In the asymptotically flat case, such modes have been identified in \cite{Richartz2014}. Here, we show that these modes exist in RNdS BHs, and that they are the limit of a new family of modes.

Comparing the NE family of modes \eqref{near extremal} to the one discussed in \cite{Cardoso:2017soq}, but initially found in \cite{Hod:2017gvn} for RN BHs, we realize that their real parts coincide, since they only depend on the choice
of BH parameters, but their imaginary parts differ slightly. In any case, as the extremal 
charge is approached, both families share the same fate; a vanishing imaginary part ($\kappa_-=
\kappa_+=0$ at extremality).
\\
\noindent{\bf{\em IV. Dominant modes and Strong Cosmic Censorship.}}
Since our purpose is to investigate the implications of charged fermions in SCC, we will restrict ourselves to choices of NE RNdS BH parameters which are problematic, since in this region $\kappa_-$ becomes comparable to the $\text{Im}(\omega)$ of the dominant QNM. For the region of interest, $\xi=1$ modes dominate all three families (see \ref{higher}).

In \cite{Ge:2018vjq} it was shown that for the choice of $\Lambda M^2=0.06$ and $Q/Q_\text{max}=0.996$ only the $\xi=1$ PS mode is relevant for SCC and
there is a region in the parameter space where $\beta>1/2$ (for $q M\lessapprox 0.53$) implying the potential violation of SCC. Quietly interesting, there was no participation of the NE modes to the determination of $\beta$ for these parameters. On the other hand, for $\Lambda M^2=0.06$ and $Q/Q_\text{max}=0.999$ a family that originates from purely imaginary modes (for $q=0$) comes into play to dominate for very small fermionic charges and quickly becomes subdominant to give its turn to the $\xi=1$ PS mode. Again, $\beta>1/2$ (for $q M\lessapprox 0.85$) so SCC may be violated.
\begin{figure*}[t]
\begin{center}
\includegraphics[scale=0.24]{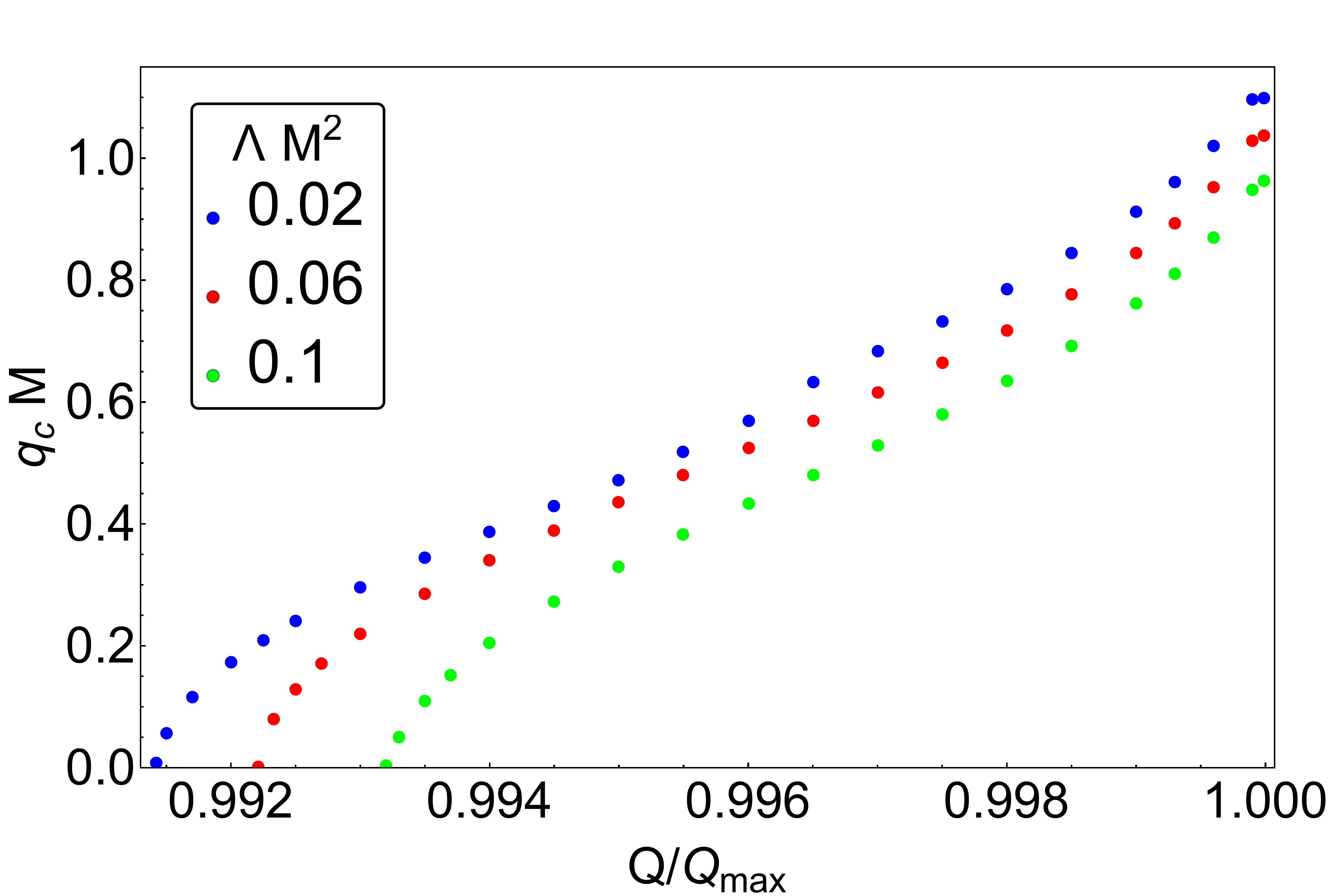}
\includegraphics[scale=0.24]{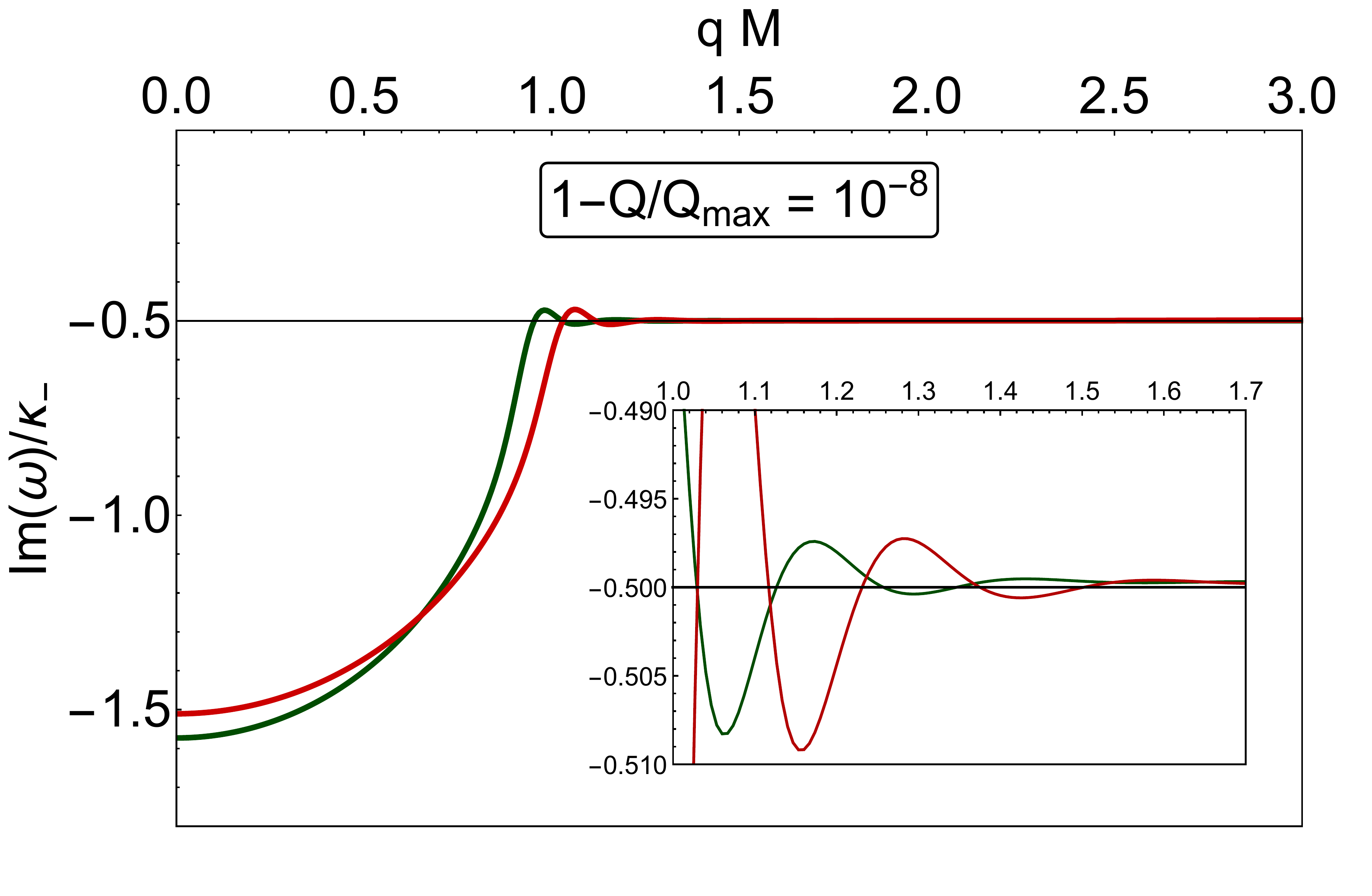}
\end{center}
\caption{{\bf Left:} Dependence of the critical fermionic charge $q_c$ on the BH charge $Q/Q_\text{max}$ and the cosmological constant $\Lambda M^2$. Here, different colors denote different choices of cosmological constants. {\bf Right:} Dominant $\xi=1$ NE QNMs of a charged, massless fermionic perturbation of a RNdS BH with $\Lambda M^2=0.005$ (red line) and $\Lambda M^2=0.06$ (green line) for $1-Q/Q_\text{max}=10^{-8}$ as a function of the fermionic charge $q M$.}
\label{qc_wiggles}
\end{figure*}

Our numerics completely agree with this picture. Here, we will be mostly interested in the case of even higher BH charges and the classification of the families originating from purely imaginary modes. To do so, we will study various choices of $\Lambda$. The BH charges we consider are:
\begin{equation}
1-Q/Q_\text{max}=10^{-3},\,\,10^{-4},\,\,10^{-5}.
\end{equation}
According to our results (see Fig. \ref{smallL}) for small BHs ($\Lambda M^2=0.005$) with $1-Q/Q_\text{max}=10^{-3}$ we see that $\beta$ is defined by the dS mode up to $q M\approx 0.5$; for larger $q$ the PS mode becomes dominant. Interestingly enough, for $q M<0.5$, the NE mode lies very close to the dS one being the first subdominant mode in this range. For larger BHs ($\Lambda M^2=0.06$) with $1-Q/Q_\text{max}=10^{-3}$ the dS mode moves rapidly to the subdominant side, giving its place to the NE mode to dominate up until $qM\approx 0.35$; for larger $q$ the PS mode dominates again. For BHs with $1-Q/Q_\text{max}\geq 10^{-4}$ the NE mode always dominate the dynamics, while the rest of the families lie out of the range of interest. 

For all cases presented, there is always a critical fermionic charge $q_c$ above which $\beta<1/2$ and SCC is preserved. In Fig. \ref{qc_wiggles} (left panel) we display the dependence of $q_c M$ on the $\Lambda M^2$ and $Q/Q_{\text{max}}$. We observe that as the BH becomes extremal a larger violation gap occurs in the parameter space. A larger $q_c M$ is also obtained for smaller cosmological constants. Similar results were obtained in \cite{PhysRevD.98.104007,Hongbao} for the case of charged scalar perturbations, although the absence of superradiance effect in fermionic fields leads to even larger regimes in the parameter space where violation of SCC may occur. 

By observing the cases with $1-Q/Q_\text{max}=10^{-5}$ we see that above $q_c$, $\beta$ lies very close to $1/2$. To examine if non-perturbative effects are present we plot $\beta$ for $\Lambda M^2=0.005,\,0.06$ and $1-Q/Q_\text{max}=10^{-8}$ versus the fermionic charge. In Fig. \ref{qc_wiggles} (right panel) we observe the existence of 
arbitrarily small oscillations of the imaginary part of the fundamental NE mode in 
highly near-extremal RNdS BHs. Such a phenomenon was previously observed for charged 
scalar perturbations in RNdS \cite{Dias:2018ufh} and gravitational perturbations in Kerr BHs \cite{Yang:2013uba}. 
These oscillations are called "wiggles" and have very small amplitude. They are suppressed exponentially fast with increasing $q$ and are precisely the non-perturbative effect that an asymptotic series, such as the WKB approximation, can easily miss, since they are highly subdominant. We believe that these wiggles were missed from the analysis of \cite{Ge:2018vjq} because they did not consider highly NE BHs.  

The ramifications of the wiggles for SCC are fierce. Our results indicate that, even for fermionic fields with $q>q_c$, there are still NE BHs for which $\beta>1/2$ and SCC may be violated, regardless of the cosmological constant, in contrast with the results in \cite{Ge:2018vjq}. Finally, notice that for all cases presented, all dominant modes originate below $\text{Im}(\omega/\kappa_-)=-1$, indicating that $\beta>1$, corresponding to a potential scenario of bounded curvature as explained in \cite{Cardoso:2017soq}.

\noindent{\bf{\em V. Conclusions.}}
We recently presented evidence in \cite{Cardoso:2017soq,PhysRevD.98.104007} for the potential failure of SCC in NE RNdS BHs under neutral and charged scalar perturbations. By utilizing (\ref{betaOld}) we performed thorough numerical analyses of $\beta$ through the calculation of QNMs of the system. Here, we extend our analysis to charged fermionic fields.

First, we provide justification that \eqref{betaOld} remains valid for charged fermionic perturbations. Then, we perform a detailed numerical computation of the dominant modes of RNdS BHs and distinguish three families of QNMs. The first family is closely related to the PS of the BH while the second is related to the existence and timescale of the dS horizon of pure dS space. The final family dominates the dynamics when NE BH charges are considered. According to our study, the only relevant region for SCC is the NE, where the surface gravity of the CH, $\kappa_-$, becomes comparable with the decay rates of the dominant QNMs. We show that all families admit their dominant modes for $\xi=1$ in this region and search for potential violation, while taking into account the entire range of $qM\geq 0$. Finally, by computing $\beta$ we consider the implications on SCC.

Our main results are shown in Figs. \ref{QNMs} - \ref{qc_wiggles} and our conclusions are summarized here. For all choices of $\Lambda M^2$ we always find a region of fermionic charges for which $\beta>1/2$ which predicts a potential failure of SCC, since the CH can be seen as singular due to the blow-up of curvature but maintain enough regularity for metric extensions to be possible beyond it. For sufficiently large fermionic charges the conjecture seems to be initially restored for highly charged RNdS BHs. After examining BHs even closer to extremality, we realize that even beyond the critical fermionic charge, violation can still occur due to the existence of wiggles.

We point out that for all cases presented, all dominant modes from the dS, PS or NE family admit $\beta>1$ for a small but significant regime of fermionic charges. This result is even more alarming for SCC since it is related to bounded curvature and therefore opens the possibility to the existence of solutions with even higher regularity across the CH. Nevertheless, if we superimpose all perturbations, then the smallest of all types of perturbations is the one relevant for SCC. Thus, the neutral scalar modes \cite{Cardoso:2017soq} admit $1/2<\beta<1$, which is enough to guarantee the blow-up of curvature components at the CH.

\noindent{\bf{\em Acknowledgments.}}
The author would like to thank Vitor Cardoso, Aron Jansen, Jo\~{a}o Costa, Peter Hintz and Hongbao Zhang for helpful discussions. The author acknowledges financial support provided under the European Union's H2020 ERC 
Consolidator Grant ``Matter and strong-field gravity: New frontiers in Einstein's theory'' grant 
agreement no. MaGRaTh--646597. 
The author would like to acknowledge networking support by the GWverse COST Action CA16104, ``Black holes, gravitational waves and fundamental physics.''
%

\begin{appendix}
\section{The definition of $\beta$ for fermions.}\label{app:beta}
In \cite{PhysRevD.98.104007} a justification of searching for $\beta>1/2$ was provided, leading to potential violation of SCC in RNdS BHs under charged scalar perturbations. Here, we prove that the same holds for charged fermions. To determine the regularity of the metric up to the CH we study the regularity of QNMs at the CH. To do so, we change to outgoing Eddington-Finkelstein coordinates which are regular there.  The outgoing Eddington-Finkelstein coordinates are obtained by replacing $t$ with $u=t-r_*$ in (\ref{RNdS_space}) to get
\begin{equation}
\label{outgoing}
ds^2=-f(r)du^2-2du dr +r^2(d\theta^2+\sin^2\theta\, d\varphi^2),
\end{equation}
with the associated electromagnetic potential
\begin{equation}
\label{outpot}
A=-\frac{Q}{r}du=-\frac{Q}{r}\left(dt-\frac{dr}{f(r)}\right).
\end{equation}
A straightforward way to write down the Dirac equation (\ref{dirac equation}) in this new coordinates is to choose a new tetrad that reproduces (\ref{outgoing}). An alternative way is to transform (\ref{dirac_2}) in the new coordinates $(u,r)$ with the associated transformed electromagnetic potential (\ref{outpot}).
\begin{figure*}[ht!]
\includegraphics[scale=0.247]{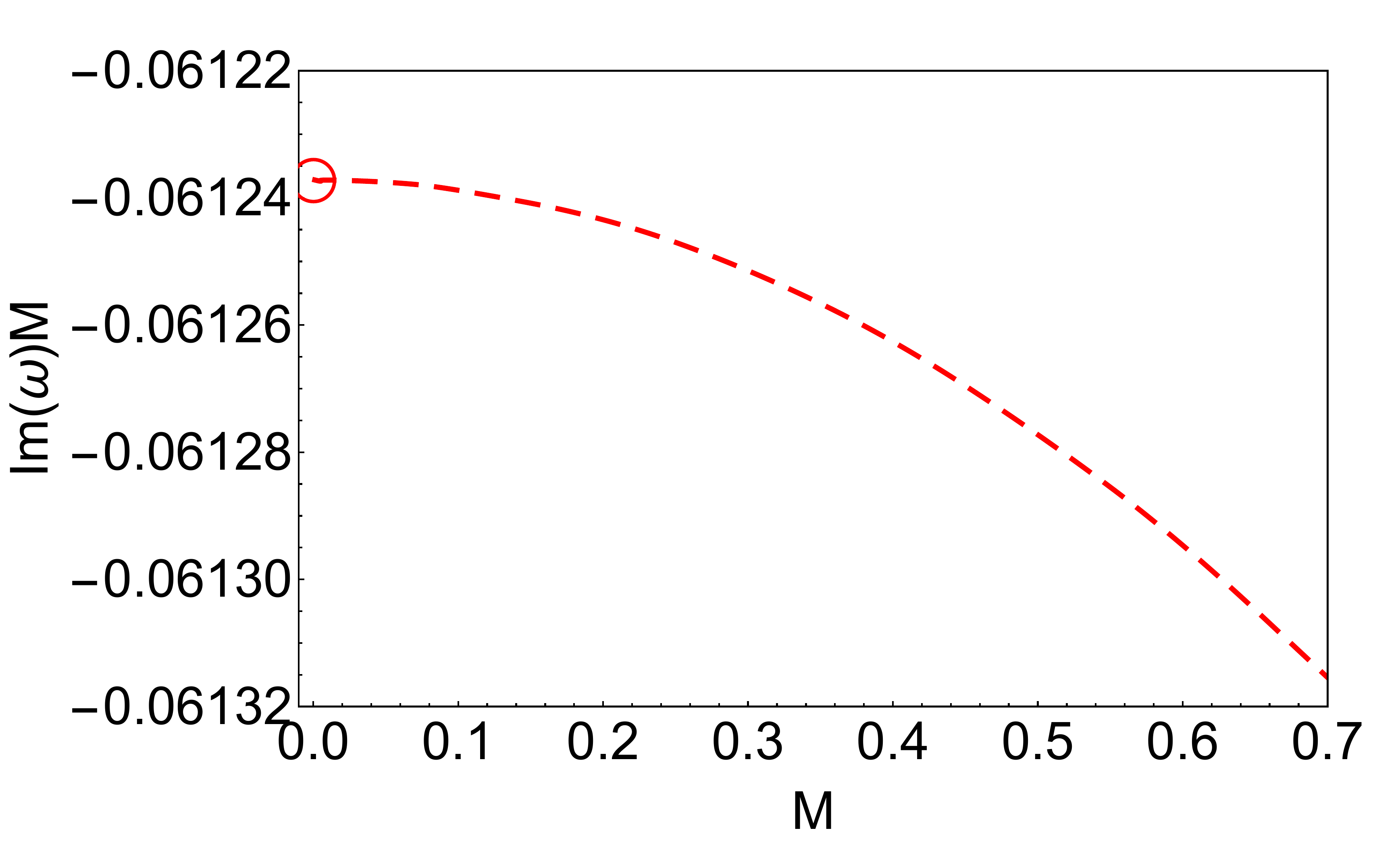}
\includegraphics[scale=0.24]{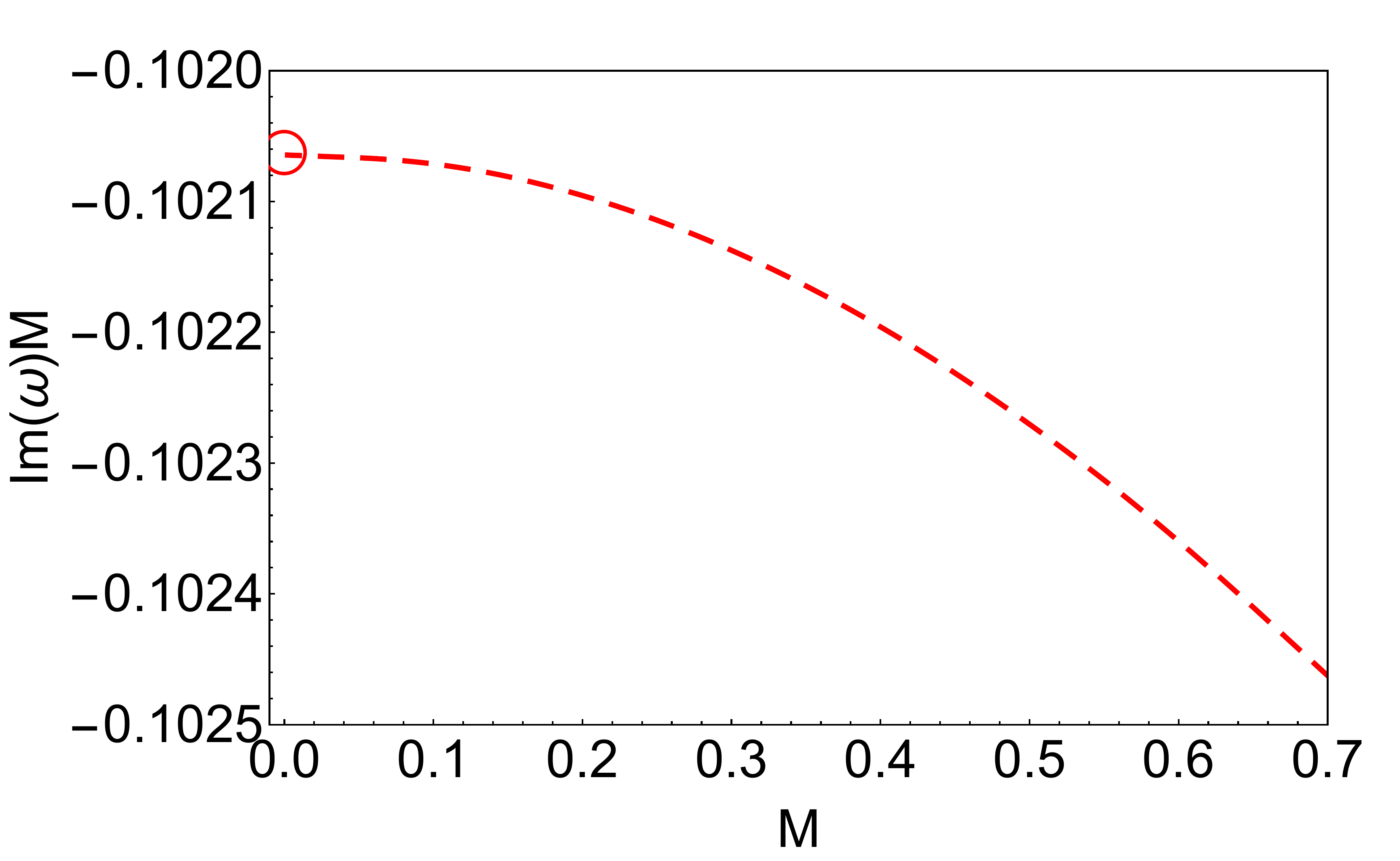}
\caption{Fundamental (left, dashed) and first overtone (right, dashed) $\xi=1$ BH dS QNM of a neutral, massless fermionic perturbation propagating on a fixed RNdS BH with $\Lambda M^2=0.005$ and $Q/M=10^{-4}$ as a function of the BH mass $M$. The red circles in each plot designate the respective pure dS QNMs.}
\label{dSconv}
\end{figure*}

To do so, we write (\ref{dirac_2}) with $m_f=0$ and perform the transformation $\partial_t=\partial_u$, $\partial_r=-\partial_u/f+\partial_r$ to obtain
\begin{align}
\nonumber
\label{dirac_3}
&\left[\frac{i\gamma^t\partial_u}{\sqrt{f(r)}}+i\sqrt{f(r)}\gamma^r\left(-\frac{\partial_u}{f}+\partial_r\right)-\frac{i\gamma^r}{r}\right. \\&
\left. +i\left(\gamma^\theta\partial_\theta+\gamma^\varphi\partial_\varphi\right)-\gamma^{t}\frac{q Q}{r\sqrt{f(r)}}+\gamma^{r}\frac{q Q}{r\sqrt{f(r)}}\right]\Psi=0.
\end{align}
By utilizing the ansatz for the Dirac wavefunctions
\begin{align}
\label{spinorout3}
\Psi^+&=e^{-i\omega u}\begin{pmatrix}
\phi^k_{j-1/2}F^+(r)\\
i\phi^k_{j+1/2}G^+(r)
\end{pmatrix},\\
\label{spinorout4}
\Psi^-&=e^{-i\omega u}\begin{pmatrix}
\phi^k_{j+1/2}F^-(r)\\
i\phi^k_{j-1/2}G^-(r)
\end{pmatrix},
\end{align}
we obtain
\begin{align}
\label{edoutf1}
f\partial_rF-\frac{\xi\sqrt{f}}{r}F+\left(\omega-\frac{qQ}{r}\right)G+i\left(\omega-\frac{qQ}{r}\right)F&=0,\\
\label{edoutf2}
f\partial_rG+\frac{\xi\sqrt{f}}{r}G-\left(\omega-\frac{q Q}{r}\right) F+i\left(\omega-\frac{qQ}{r}\right)G&=0,
\end{align}
where $\xi=\pm 1,\pm 2,\dots$. By solving with respect to $F$ we obtain
\begin{align}
\nonumber
\label{int}
-\frac{2\xi \omega f^{3/2}}{r(qQ-r\omega)} F-\frac{\xi\sqrt{f}}{r}f^\prime F-2\frac{\xi^2 f}{r^2}F+\frac{2qQf}{r(qQ-r\omega)}f\partial_r F\\
-\frac{4iqQ}{r}f\partial_r F+4i\omega f\partial_rF+2(f\partial_r)^2F=0.
\end{align}
It can be shown that the mode solutions of (\ref{int}) are conormal at $r=r_-$, meaning that they grow at the same rate $|r-r_-|^\lambda$. Thus, if $F\sim|r-r_-|^\lambda$ then the first four terms have higher regularity than the rest, where $f \sim |r-r_-|$ near the CH modulo irrelevant terms. This means that these terms can be neglected, which leads to a regular-singular ordinary differential equation near $r=r_-$ of the form $PF=0$ with the operator\footnote{The same operator arises for the field $G$ with $PG=0$ by following exactly the same steps.}
\begin{equation}
\label{operator}
P=(f\partial_r)^2+2i\omega(f\partial_r)-\frac{2iq\omega}{r}(f\partial_r).
\end{equation}

It is convenient to use $f$ as a radial coordinate instead of $r$, so $\partial_r=f^\prime\partial_f=f^\prime(r_-)\partial_f$ near the CH modulo irrelevant terms. Moreover, the surface gravity at the CH is $\kappa_-=-f^\prime(r_-)/2$ so $f\partial_r=-2\kappa_-(f\partial_f)$. Thus, (\ref{operator}) becomes
\begin{align}
\label{indicial}
\nonumber
\frac{{P}}{4\kappa_-^2}&=(f\partial_f)^2-\frac{i\omega}{\kappa_-}(f\partial_f)+\frac{iqQ}{\kappa_-r_-}(f\partial_f)\\
&=f\partial_f\left(f\partial_f-\left(\frac{i\omega}{\kappa_-}-\frac{iqQ}{\kappa_-r_-}\right)\right).
\end{align}
It remains to calculate the allowed growth rates $\lambda$, i.e. indicial roots of the differential operator (\ref{indicial}). Acting with $|f|^\lambda$ we get
\begin{equation}
\label{polynomial}
\frac{{P}}{4\kappa_-^2}|f|^\lambda=\lambda\left(\lambda-\left(\frac{i\omega}{\kappa_-}-\frac{iqQ}{\kappa_-r_-}\right)\right)|f|^\lambda,
\end{equation}
The indicial roots are the roots of the quadratic polynomial (\ref{polynomial}), namely
\begin{equation}
\lambda_1=0,\,\,\,\,\,\,\,\,\,\,\,\,\,\lambda_2=\frac{i\omega}{\kappa_-}-\frac{iqQ}{\kappa_-r_-}.
\end{equation}
The root $\lambda_1=0$ corresponds to mode solutions which are approximately constant, i.e. remain smooth at the CH and are irrelevant for SCC, while $\lambda_2$ corresponds to asymptotics
\begin{equation}
|f|^{\lambda_2}\sim|r-r_-|^{\frac{i\omega}{\kappa_-}}|r-r_-|^{-\frac{iqQ}{\kappa_-r_-}}.
\end{equation}
If we consider QNMs of the form $\omega=\omega_R-i\omega_I$ then
\begin{equation}
|f|^{\lambda_2}\sim|r-r_-|^\frac{\omega_I}{\kappa_-}|r-r_-|^{i\left(\frac{\omega_R}{\kappa_-}-\frac{qQ}{\kappa_-r_-}\right)}.
\end{equation}
The second factor is purely oscillatory, so the only relevant factor for SCC is $|r-r_-|^\frac{\alpha}{\kappa_-}$ with $\alpha:=-\text{Im}{\omega}$ the spectral gap define in \cite{Cardoso:2017soq}. This function lies in the Sobolev space $H^s$ for all $s<\frac{1}{2}+\frac{\alpha}{\kappa_-}$. 

Considering scalar perturbations, we require locally
square integrable gradient of the scalar field at the CH, i.e., the mode solutions should belong to the Sobolev space $H^1_\text{loc}$ for the Einstein's field equations to be satisfied weakly at the CH. 

In our case, the Einstein-Hilbert stress-energy tensor of the fermionic field lying on the right-hand-side of Einstein's equations has the form \cite{Toth:2015cda}
\begin{align}
\label{tmn}
\nonumber
T^{\mu\nu}=&\frac{1}{4}\left(\bar{\Psi}i\gamma^\mu\left(\nabla^\nu-iqA^\nu\right)\Psi+\bar{\Psi}i\gamma^\nu\left(\nabla^\mu-iqA^\mu\right)\Psi\right.\\
&\left.-\left(\nabla^\mu+iqA^\mu\right)\bar{\Psi}i\gamma^\nu\Psi-\left(\nabla^\nu+iqA^\nu\right)\bar{\Psi}i\gamma^\mu\Psi\right)
\end{align}
and again this leads to the requirement of square integrability of the gradient of the fermionic field. Thus, the mode solutions should belong to the Sobolev space $H^1_\text{loc}$ for our metric to make sense as a weak solution of Einstein's field equations at the CH. This provides the justification for our search for BH parameters for which $\beta>1/2$.

\begin{figure*}[t]
\includegraphics[scale=0.295]{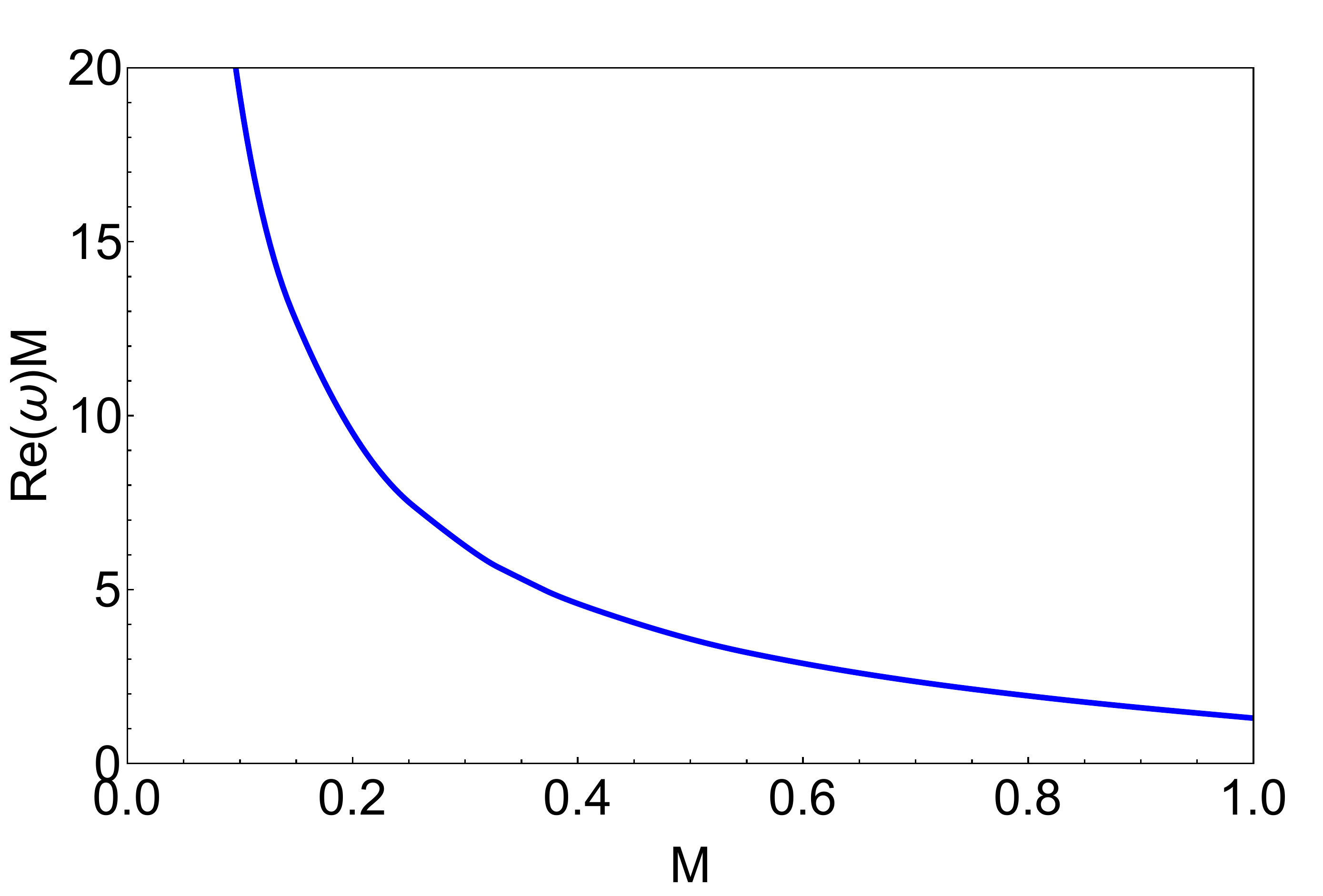}
\includegraphics[scale=0.3]{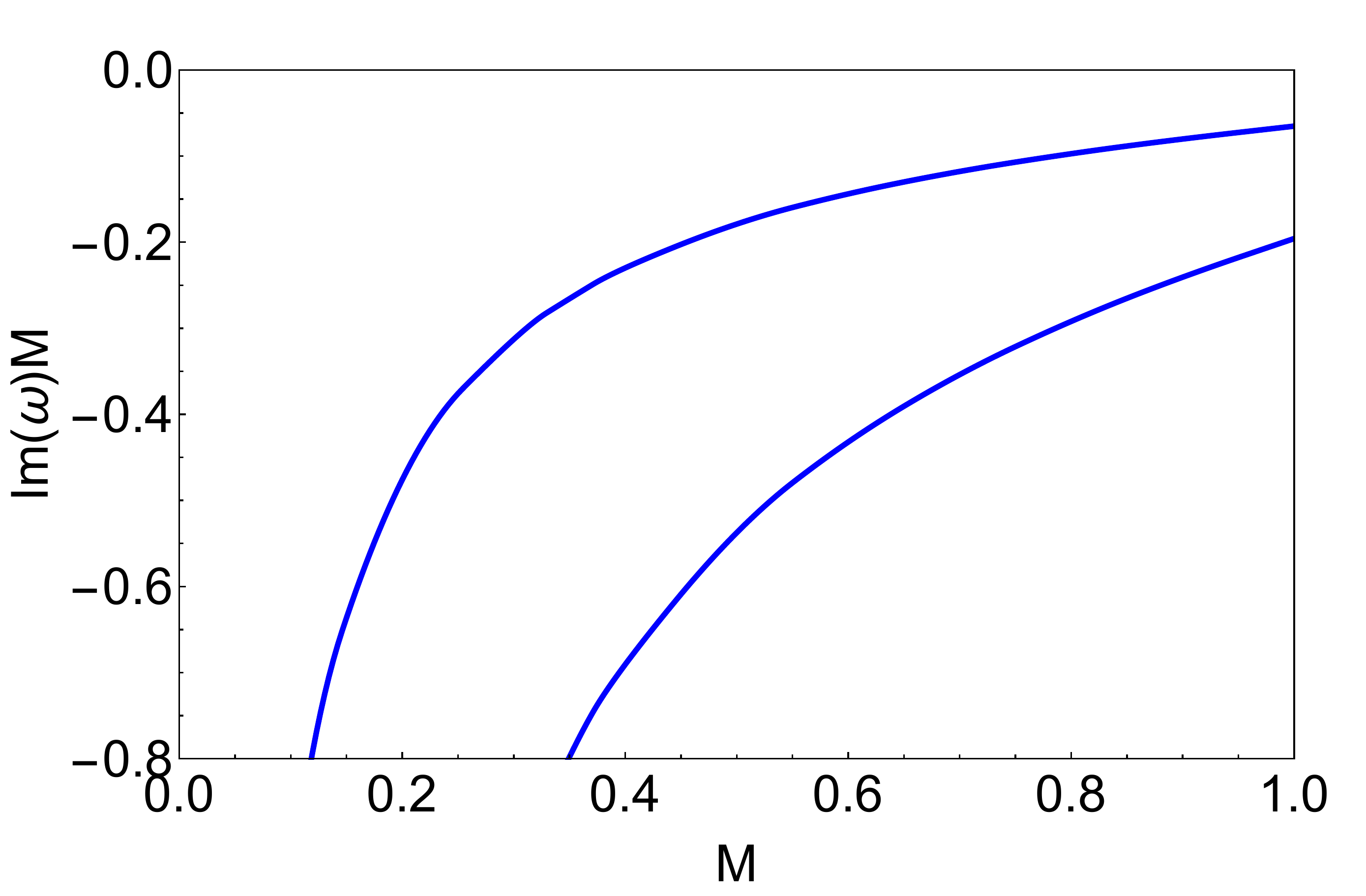}
\caption{Real (left) and imaginary (right) part of the $\xi=10$ fundamental and first overtone PS QNM of a neutral, massless fermionic perturbation propagating on a fixed RNdS BH with $\Lambda M^2=0.06$ and $Q/M=10^{-4}$ as a function of the BH mass $M$. The real parts coincide in the range shown.}
\label{PSconv}
\end{figure*}

\section{Higher $\xi$ modes}{\label{higher}}
In this appendix we verify the expectation that the higher $\xi$ QNMs do not affect Strong Cosmic Censorship. In Fig. \ref{higher_k} we show the $\xi=1,2,3$ modes. The ones depicted with blue colors belong to the PS family, since they originate from complex modes for $q=0$ and follow their pattern. The ones depicted with green colors belong to the NE family, since they originate from purely imaginary modes for $q=0$ and follow their pattern. The dS modes are not present in the range of interest since they are too subdominant for the chosen cosmological constant. We clearly see that the modes defining $\beta$ according to (\ref{betaOld}) will be the $\xi=1$ QNMs. The same holds for other choices of $\Lambda$.

For completeness, in Table \ref{table} we show various modes from different families with $\xi=1,10$ for various choices of $Q$, $q$ and $\Lambda$. We compare the neutral $\xi=10$ PS modes with a WKB approximation for arbitrarily large $\xi$ and verify that indeed the imaginary parts lie very close. It is apparent that for NE charges $\xi=1$ modes always dominate. It is also apparent that the only way for $\xi\rightarrow\infty$ modes to be the dominant ones of the PS family is for very small cosmological constants. Specifically, for $\Lambda M^2=0.001$, $\xi\rightarrow\infty$ modes are dominant up to a critical BH charge $Q_c/M\approx 0.866$. Above $Q_c/M$, $\xi=1$ modes dominate the PS family. E.g. for $Q/M=0.865$ and $q=0$ the dominant ($\xi\rightarrow\infty$) PS mode admits $\text{Im}(\omega_\text{PS})/\kappa_-=-0.0479$, while the dominant ($\xi=1$) dS mode admits $\text{Im}(\omega_\text{dS})/\kappa_-=-0.0135$; the NE family is too subdominant for this BH charge. None of those modes can potentially violate SCC so it becomes a necessity to search closer to extremality, where we are aware that $\kappa_-$ becomes comparable to $\text{Im}(\omega)$.\footnote{For the case presented, the surface gravity is $\sim 100$ larger than the decay rate of the dominant mode.} Since $\beta$ is maximal at $q=0$, any $qM>0$ will make $\text{Im}(\omega)/\kappa_-$ even smaller. Finally, for larger $\Lambda$, $Q_c$ decreases, moving even further away from extremality.

Considering the above, we are convinced that throughout the parameter space in study, $\xi=1$ indeed gives the dominant modes for all families.

\begin{table}[h]
\centering
\scalebox{0.8}{
\begin{tabular}{||c| c | c ||} 
\hline
  \multicolumn{3}{||c||}{$Q/M=10^{-1}$} \\
   \hline
\hline
  \multicolumn{3}{||c||}{$\Lambda M^2=0.005$} \\
   \hline
    $\xi$ & $qM=0$ & $qM=0.1$ \\ [0.5ex] 
   \hline
   1 &  $\omega_\text{PS}$= 0.1795 - 0.0947 i & $\omega_\text{PS}$= -0.1760 - 0.0941 i   \\  
   & $\omega_\text{dS}$= -0.0614 i & $\omega_\text{dS}$= -0.00003 - 0.0614 i \\
   \hline
   10 & $\omega_\text{PS}$= 1.8831 - 0.0941 i& $\omega_\text{PS}$= -1.8797 - 0.0940 i \\ 
   \hline
   WKB &$\,\,\,\,\,\,\,\,\,\,\,\,\,\,\,\,\,\,\,\,\,\,\,\,\,\,\,\,\,\,\,\,\,\,\,\,$- 0.0941 i & - \\
   \hline\hline
  \multicolumn{3}{||c||}{$\Lambda M^2=0.06$} \\
   \hline
    $\xi$ & $qM=0$ & $qM=0.1$ \\ [0.5ex] 
   \hline
      1 & $\omega_\text{PS}$= 0.1280 - 0.0650 i & $\omega_\text{PS}$= -0.1247 - 0.0647 i    \\  
      & $\omega_\text{dS}$= -0.2170 i & $\omega_\text{dS}$= -0.0003 - 0.2170 i \\
      \hline
      10 &$\omega_\text{PS}$= 1.3097 - 0.0654 i & $\omega_\text{PS}$= -1.3064 - 0.0654 i \\ 
      \hline
      WKB &$\,\,\,\,\,\,\,\,\,\,\,\,\,\,\,\,\,\,\,\,\,\,\,\,\,\,\,\,\,\,\,\,\,\,\,$ - 0.0654 i & - \\
      \hline
\end{tabular}
}
\scalebox{0.833}{
\begin{tabular}{||c| c | c ||} 
\hline
  \multicolumn{3}{||c||}{$Q/Q_\text{max}=1-10^{-3}$} \\
   \hline
\hline
  \multicolumn{3}{||c||}{$\Lambda M^2=0.005$} \\
   \hline
    $\xi$ & $qM=0$ & $qM=0.1$ \\ [0.5ex] 
   \hline
    & $\omega_\text{PS}$=0.2353 - 0.0865 i &   $\omega_\text{PS}$=-0.1875 - 0.0852 i  \\  
   1& $\omega_\text{dS}$=-0.0613 i & $\omega_\text{dS}$=-0.0003 - 0.0612 i \\
   & $\omega_\text{NE}$=-0.0671 i & $\omega_\text{NE}$=0.1004 - 0.0669 i\\
   \hline
   10 & $\omega_\text{PS}$=2.4650 - 0.0872 i & $\omega_\text{PS}$=-2.4152 - 0.0872 i \\ 
   \hline
   WKB &$\,\,\,\,\,\,\,\,\,\,\,\,\,\,\,\,\,\,\,\,\,\,\,\,\,\,\,\,\,\,\,\,$ - 0.0870 i& - \\
   \hline\hline
  \multicolumn{3}{||c||}{$\Lambda M^2=0.06$} \\
   \hline
    $\xi$ & $qM=0$ & $qM=0.1$ \\ [0.5ex] 
   \hline
      1 & $\omega_\text{PS}$=0.2016 - 0.0708 i & $\omega_\text{PS}$=0.2548 - 0.0692 i    \\  
      & $\omega_\text{NE}$=-0.0611 i & $\omega_\text{NE}$=0.0974 - 0.0609 i \\
      \hline
      10 &$\omega_\text{PS}$=2.0918 - 0.0716 i & $\omega_\text{PS}$=2.1436 - 0.0715 i \\ 
      \hline
      WKB & $\,\,\,\,\,\,\,\,\,\,\,\,\,\,\,\,\,\,\,\,\,\,\,\,\,\,\,\,\,\,\,\,\,\,$- 0.0718 i& - \\
      \hline
\end{tabular}
}
\caption{Lowest lying fermionic QNMs of RNdS BH for various $Q$, $q$, $\Lambda$ and $\xi$.}
\label{table}
\end{table}
\section{Convergence of the families}\label{dS}
In this appendix we demonstrate the convergence of the BH dS modes to the pure dS QNMs, as well as the behavior of the PS modes, for vanishing $M$. In Fig. \ref{dSconv} we plot with dashed lines the fundamental ($n=0$) and first overtone ($n=1$) BH dS modes and illustrate that as the BH disappears ($M\rightarrow 0$) the QNMs tend smoothly to the exact pure dS fermionic QNMs which are denoted with red circles. Here, we only demonstrate this for a specific choice of parameters but our numerics reveal that the smooth convergence happens for all BH parameters. We realize that the increment of the BH mass affects weakly the dS family. It is important to note that equivalent results for the BH dS family were obtained in RNdS under scalar perturbations \cite{Cardoso:2017soq}.

The story is different for the PS family of modes (see Fig. \ref{PSconv}). As $M\rightarrow 0$ the real and imaginary parts diverge. This occurs due to the shrinking of the photon sphere. If we consider, for example, a perturbed string with a specific length that vibrates, then by continuously decreasing its length we will observe that the vibrations will have increasingly higher frequency and smaller timescales until the point where the sting vanished and oscillations cease to exist. The same happens for a dissipative BH like the one in study as it disappears. 

\end{appendix}
\FloatBarrier
\bibliography{references}

\end{document}